\documentclass[%
 reprint,
superscriptaddress,
 amsmath,amssymb,
 aps,
floatfix,
]{revtex4-2}

\usepackage{color}
\usepackage{graphicx}
\usepackage{dcolumn}
\usepackage{bm}
\usepackage{hyperref}
\usepackage{natbib}
\usepackage{comment}
\usepackage{ulem}
\usepackage{float}

\usepackage[makeroom]{cancel}

\begin{document}

\title{Light-driven interlayer propagation of collective-mode excitations \texorpdfstring{\\}{} in layered superconductors}

\author{Niklas Ziereis}
\affiliation{%
Department of Physics, University of Tokyo, 7-3-1 Hongo, Tokyo 113-0033, Japan}
\affiliation{%
Arnold Sommerfeld Center, Ludwig-Maximilians University, 80333 Munich, Germany}%
\affiliation{Technical University of Munich, TUM School of Natural Sciences, Physics Department, 85748 Garching, Germany}
\author{Kazuaki Takasan}
\affiliation{%
Department of Physics, University of Tokyo, 7-3-1 Hongo, Tokyo 113-0033, Japan}%

\author{Naoto Tsuji}
\affiliation{%
Department of Physics, University of Tokyo, 7-3-1 Hongo, Tokyo 113-0033, Japan}
\affiliation{%
RIKEN Center for Emergent Matter Science, Wako, Saitama 351-0198, Japan}%

\date{\today}
             
\begin{abstract}
Superconductors exhibit a nonlinear interaction with an applied light, which can resonantly excite the collective amplitude (Higgs) mode. Here we study light-induced dynamics of layered superconductors, where each layer is coupled to adjacent layers via the Josephson coupling and the first few layers near the surface are driven by an in-plane-polarized light. We study the system under the assumption that the interlayer Coulomb interactions are sufficiently screened out and that the phase-difference mode becomes available in the low-energy regime. We find that interlayer transport is induced via excitations of the collective amplitude and phase-difference modes, even when the applied electric field is parallel to the planes. We provide analytic calculations as well as numerical simulations of the real-time dynamics, and investigate the influence on the light-induced interlayer Josephson current and intralayer third-harmonic generation.
\end{abstract}
\maketitle
\section{Introduction}
\label{sec:intro}
\begin{figure}
    \centering
    \includegraphics{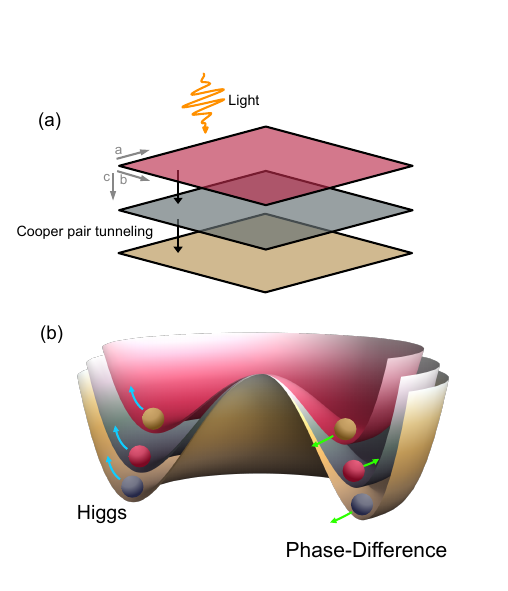}
    \caption{\textbf{(a)} Schematic picture of a Josephson-coupled three-layer superconductor driven by light. \textbf{(b)} Free energy landscape of a three-layer system: A Mexican-hat potential is associated with each superconducting layer. 
    The phase modes of the individual layers combine to form two phase-difference modes and one global phase mode with all the phases oscillating in unison. The latter one is the Nambu-Goldstone mode of the layered system which is pushed to the plasma frequency by the Anderson-Higgs mechanism. What remain at low energies are the phase-difference modes (one of which is indicated by the green arrows). There are also Higgs amplitude modes for each layer (shown by the blue arrows).}
    \label{fig:layers}
\end{figure}
Ever since the discovery of high-$T_c$ superconductivity in cuprates, the studies of layered superconductors have attracted the interest of many researchers \cite{klemm_layered_2011,dienst_bi-directional_2011,fertig_collective_1991,artemenko_collective_1995,laplace_josephson_2016,fausti_light-induced_2011}. Their highly anisotropic superconducting and normal state properties are well described by intrinsic Josephson junctions (IJJ) \cite{laplace_josephson_2016}, i.e., two-dimensional superconducting layers of CuO$_2$ separated by insulating barriers. \\
\indent Many of the interesting physical effects in these systems arise from the anomalous Josephson current \cite{josephson_possible_1962, josephson_coupled_1964, ambegaokar_tunneling_1963}, which is associated with the tunneling of Cooper pairs, rather than quasiparticles \cite{mahan_many-particle_2000}. Among these is the occurrence of a plasma mode normal to the superconducting layers \cite{gabriele_generalized_2022, okamoto_theory_2016,bulaevskii_time-dependent_1994,savelev_terahertz_2010,fertig_collective_1991, artemenko_collective_1995}, that is typically in the GHz to THz regime and can be detected by a sharp drop in the c-axis reflectivity of the materials above a characteristic energy $\omega_J$ (Josephson plasma edge). This gap arises, because in charged superconductors, Coulomb interactions give rise to a mass of the otherwise gapless phase mode \cite{ohashi_goldstone_1997} that couples to the gauge field. Remarkably, one can retrieve the acoustic dispersion of the Goldstone mode near $T_c$, where the large number of normal electrons completely screen out interlayer Coulomb interactions \cite{carlson_superconducting_1973, wong_effects_1988, ohashi_goldstone_1997} (Carlson-Goldman mode). \\
\indent Below $\omega_J$ the excitation of surface plasma modes, which arise as solutions to the interlayer sine-Gordon equation with imaginary frequency \cite{bulaevskii_time-dependent_1994} has been considered theoretically \cite{savelev_surface_2005, lu_surface_2020}. However, exciting these modes experimentally is far from trivial, making the use of atomic force microscope tips or total internal reflection of prisms \cite{laplace_josephson_2016} necessary. On the other hand, it is possible to optically induce interlayer superconducting transport by applying c-axis polarized light with high field transients of tens of $\mathrm{kV cm^{-1}}$ which induces interlayer voltage drops \cite{dienst_bi-directional_2011}.

In this way, past studies on optically induced c-axis transport in layered superconductors have focused on c-axis polarized light. In contrast, to our knowledge, the possibility of c-axis transport induced by in-plane polarized light has not been well explored to this date. We propose a mechanism resulting in the propagation of collective excitations in the direction normal to the layers using the in-plane ploarized light, based on the Anderson's pseudospin picture \cite{anderson_random-phase_1958}.

It is known that the dynamics of superconductors can be effectively mapped to that of spin systems (Anderson's pseudospins) within the time-dependent Bardeen-Cooper-Schrieffer (BCS) mean-field theory, which provides a useful way to represent and visualize a time-evolving superconducting state. In clean systems, pseudospins are defined in the momentum space, where the $z$ component of pseudospins corresponds to the occupation of electrons, while the $x$ and $y$ components represent the density of Cooper pairs for each momentum \cite{anderson_random-phase_1958}. This formalism is well-suited to describe the quadratic interaction with an ac electric field, since it corresponds to a coherent precession of the pseudospins in an oscillating pseudomagnetic field (i.e., an effective magnetic field acting on pseudospins) \cite{tsuji_theory_2015, matsunaga_light-induced_2014, murotani_theory_2017,cea_nonlinear_2016,tsuji_higgs_2024}. If an ac field with frequency $\Omega$ is applied to a superconductor, the superconducting order parameter exhibits oscillations with frequency $2\Omega$, which has been shown to resonate with collective amplitude and phase modes \cite{tsuji_theory_2015,murotani_theory_2017,matsunaga_light-induced_2014, shimano_higgs_2020, li_amplitude_2023, li_collective_2024}.

Let us consider the application of an ab-axis-polarized light to a layered structure (see Fig. \ref{fig:layers}(a)). If the thickness of a single layer is at least of the order of the London penetration depth (as can be the case for systems of artificial stacks of Josephson junctions \cite{klemm_layered_2011}), pseudospin precession is expected to be confined to the first few layers. However, the precessing pseudospins generate a pseudomagnetic field that also acts on adjacent layers, even if they are not directly affected by the incident light. This will induce pseudospin precession throughout the entire layered structure, indicating a highly nonlocal nature of this phenomenon. We will find that the application of light to the first layer is sufficient to induce resonant excitation of the collective Higgs amplitude mode in all the layers. Similarly, a phase-difference mode can be excited to propagate along the c-axis, despite the applied ac-field light being polarized in the ab-plane.

This paper is organized as follows: In Sec. \ref{sec:2}, we will study a phenomenological Ginzburg-Landau (GL) theory of layered superconductors to gain intuition for the structure of the collective modes in coupled superconductors. In Sec. \ref{sec:3} we present a minimal microscopic model of Josephson coupled layered superconductors based on the BCS mean-field Hamiltonian and derive linearized equations of motion for the amplitude and phase of the superconducting order parameters based on a formalism developed in \cite{tsuji_theory_2015,murotani_theory_2017}. In Sec. \ref{sec:4} we subsequently study analytic solutions that describe the nonlinear excitations of collective modes, and compare these results with numerical simulations. Section \ref{sec:5} is then devoted to the two main observables, the nonlinear ab-current that results in third-harmonic-generation and the interlayer Josephson current.

Throughout the paper, we will use the natural units and set the lattice constant $a=1$.
\section{Phenomenological Ginzburg-Landau Theory}
\label{sec:2}
We will start by discussing a phenomenological GL-theory of Josephson coupled layered superconductors, which provides a good intuition about the collective modes. A schematic picture of the GL free energy is shown in Fig. \ref{fig:layers}(b) for the case of three layers, but the generalization to \textit{N} layers is straightforward. The free energy of each of the \textit{N} layers is described by a Mexican-hat potential. Fluctuations around the minimum give rise to \textit{N} independent Higgs amplitude modes (one for each layer). Additionally, there are \textit{N} phase degrees of freedom, one of which is the Nambu-Goldstone mode for the total U(1) phase that gets pushed to the plasma frequency by the Anderson-Higgs mechanism. The remaining $N-1$ modes correspond to oscillations of the phase difference.

In the rest of this section, we will restrict ourselves to a bilayer system for simplicity. Nevertheless, this model illustrates the essential features that were introduced in the preceding heuristic discussion. The GL free energy density for the superconducting order parameter $\Psi$ of two Josephson-coupled superconducting layers with inversion symmetry in the presence of a gauge field $\mathbf{A}(\mathbf{r})$ can be written down as \cite{murray_large_2010, tsuji_theory_2015,kamatani_optical_2022}:
\begin{align}
    &\mathcal{F} = \sum_{i=1,2}\mathcal{F}_i + \mathcal{F}_{1,2} ,
\end{align}
where the free energy density of an individual layer is given by:
\begin{align}
\mathcal{F}_i=&c_1|\Psi_i(\boldsymbol{r})|^2  +\frac{c_2}{2} |\Psi_i(\boldsymbol{r})|^4 + 
\frac{1}{2m^*}|\mathcal{D}_i\Psi_i(\boldsymbol{r})|^2 
\end{align}
and the Josephson-coupling  between the two layers is described by:
\begin{align}
&\mathcal{F}_{1,2}=
\epsilon\Psi_1(\boldsymbol{r})\Psi_2^*(\boldsymbol{r}) + \epsilon_1\mathcal{D}_1^*\Psi^*_1(\boldsymbol{r})\mathcal{D}_2\Psi_2(\boldsymbol{r})  
\end{align}
Here, $\bm r$ labels the two-dimensional coordinate in the ab-plane, while the index $i\in \{1,2\}$ is used to distinguish between individual layers and $\mathcal{D}_i = -i\nabla  -e^*\mathbf{A}_i(\mathbf{r})$. $e^*=2e$ is the charge of a Cooper pair, $m^*$ is its mass, $\mathbf{A}_i(\mathbf{r})$ and $\phi_i(\mathbf{r})$ are the electromagnetic vector and scalar potential on the $i$-th layer, which we assume to be polarized in the ab-plane, $c_1$ and $c_2$ are coupling constants. We take a polar decomposition, $\Psi_i = (\Psi_0^i + H_i(\mathbf{r}))e^{i\theta_i(\mathbf{r})}$, where $\Psi_0^i$ denotes the superconducting order parameter in equilibrium and $H_i(\mathbf{r}), \theta_i(\mathbf{r}) \in \mathbb{R}$ describe fluctuations of the amplitude (Higgs) and phase (Nambu-Goldstone) modes for each layer.  In the following, we will suppress the dependence on $\bm r$ for notational convenience.

Let us first consider the case for $\epsilon=\epsilon_1=0$, where the two layers are decoupled. We begin by expanding the single-layer free energy $\mathcal{F}_i$  in terms of the fluctuations $H_i$ and $\theta_i$, which results in
\begin{align}
    \mathcal{F}_i &=2c_1 H_i^2 -\frac{1}{2m^*}(\nabla H_i)^2  \\
    &- \frac{e^{*2}}{2m^*}\left(\mathbf{A}_i - \frac{1}{e^*}\nabla\theta_i\right)^2\left(\Psi_0^i+H_i\right)^2 + \ldots .\nonumber
\end{align}
We observe that the free energy density $\mathcal{F}$ is invariant under the local gauge transformations: $\Psi_i(\mathbf{r}) \to e^{ie^*\chi(\mathbf{r})}\Psi_i(\mathbf{r})$ and $ \mathbf{A}_i(\mathbf{r})\to\mathbf{A}_i(\mathbf{r}) + \nabla\chi(\mathbf{r})$.
Hence, we can gauge out $\theta_i$ in the combination $\mathbf{A}_i - \frac{1}{e^*}\nabla\theta_i$ and thus remove the phase degree of freedom. This results in the Anderson-Higgs mechanism, where the Nambu-Goldstone mode is pushed to the plasma frequency. After the gauge transformation, $\mathcal{F}_i$ takes the following form to the first order in the fluctuations:
\begin{align}
    \mathcal{F}_i= - \frac{e^{*2}\Psi_0^{i2}}{2m^*}\mathbf{A}_i^2 + \frac{e^{*2}\Psi_0^i}{m*}\mathbf{A}_i^2H_i + \mathcal{O}(H_i^2,\theta_i^2,H_i\theta_i).
\end{align}
The first term describes the mass of the gauge field that is generated by the Anderson-Higgs mechanism. The second term results in a nonlinear excitation of the Higgs mode through the squared electromagnetic field \cite{tsuji_theory_2015}, which can microscopically induce precession of Anderson's pseudospins.

We now turn to the case of Josephson-coupled superconductors with $\epsilon, \epsilon_1 \neq 0$. Specifically, we define $\mathcal{F}_{\epsilon_1}$ and $\mathcal{F}_{\epsilon_2}$ to be the terms proportional to $\epsilon_1$, which describes interactions with the electromagnetic field. Once again, expanding in terms of $\theta_i$ and $H_i$ yields:
\begin{align}
    \mathcal{F}_{\epsilon_1} =& 2e^{*2}\Big[ \Re(\epsilon_1)\left(\mathbf{A}_1-\frac{1}{e^{*}}\nabla\theta_1\right)\left(\mathbf{A}_2-\frac{1}{e^{*}}\nabla\theta_2\right)\nonumber\\
    +&\Im(\epsilon_1)\theta_{\mathrm{rel}}\mathbf{A}_1\mathbf{A}_2\Big]\times \left(\Psi_0^2+H_2\right)\left(\Psi_0^1+H_1\right) + \ldots ,
\end{align}
where we defined the relative phase $\theta_{\mathrm{rel}}=\theta_1-\theta_2$. While the Nambu-Goldstone modes of the individual layers can be gauged out in the first term, their difference, $\theta_{\mathrm{rel}}$ cannot be removed simultaneously and thus survives the Anderson-Higgs mechanism. This mode is known as the Leggett mode in the studies of multiband superconductors \cite{leggett_number-phase_1966}
and becomes the Josephson plasma mode in layered superconductors with $N\gg2$ layers. This mode is not subject to the Anderson-Higgs mechanism, as it does not constitute a Goldstone mode, generated by U(1) symmetry. It is therefore not pushed to the plasma frequency of the individual layers, but a gap opens up in the presence of the screened Coulomb repulsion. Similarly to the case of the Higgs mode, the phase-difference mode may be excited nonlinearly by the squared vector potential \cite{murotani_theory_2017}.

Now we consider the case, where the thickness of the first layer exceeds the London penetration depth.
We first perform a gauge transformation with $\chi(\mathbf{r})=\frac{1}{e^*}\theta_1$.
before assuming that $\mathbf{A}_2 \approx 0$. In this case we have
\begin{equation}
F_{\epsilon_1}=2e^*\Im(\epsilon_1)\theta_{\mathrm{rel}}\mathbf{A}_1\nabla\theta_1 +\mathcal{O}(H_i^2,\theta_i^2,H_i\theta_i).
\end{equation}
From the equations of motion for $\theta_1$ it is apparent, that $\nabla \theta_1 = e^*\mathbf{A}_1 +\ldots$. Thus we have that
\begin{equation}
F_{\epsilon_1}=2e^*\Im(\epsilon_1)\theta_{\mathrm{rel}}\mathbf{A}_1^2 + \ldots,
\end{equation}
which demonstrates that the relative phase mode $\theta_{\mathrm{rel}}$ will be nonlinearly excited and oscillate with twice the frequency of the applied light. The oscillating phase difference will result in an oscillating Josephson current $I_J=I_c\sin(\theta_{\mathrm{rel}})$ propagating along the c-axis between the layers, induced by ab-polarized light.

\section{\label{sec:3}Microscopic Model}
After discussing the GL theory, we turn to a microscopic model based on the BCS mean-field Hamiltonian.
We start from a pairing Hamiltonian describing a stack of $N$ $s$-wave superconductors separated by insulating barriers and each interacting with its adjacent layers via the tunneling of Cooper pairs:
\begin{eqnarray}
\mathcal{H}_N &&= 
\sum_{\mathbf{k}, \sigma, n}\epsilon_{n,\mathbf{k}-e\mathbf{A}_n(t)}c^\dagger_{\mathbf{k}\sigma, n}c_{\mathbf{k}\sigma ,n}  \nonumber \\
&&+\sum_{\mathbf{k}, \mathbf{k}', n} V_{n,n} c^\dagger_{\mathbf{k}\uparrow ,n}c^\dagger_{-\mathbf{k}\downarrow ,n}c_{-\mathbf{k}'\downarrow ,n} c_{\mathbf{k}'\uparrow ,n} \label{eq: Hamiltonian}\\
&&+\sum_{\mathbf{k,k'},n}V_{n,n+1}c^\dagger_{\mathbf{k}\uparrow, n+1}c^\dagger_{-\mathbf{k}\downarrow ,n+1}c_{-\mathbf{k}'\downarrow, n}c_{\mathbf{k}'\uparrow ,n} + \mathrm{h.c.}\nonumber ,
\end{eqnarray}
where $n\in\left\{1,...,N\right\}$ labels the $n$th superconducting layer, $c^\dagger_{\mathbf{k} \sigma n}$ creates an electron with momentum $\mathbf{k}$ and energy $\epsilon_{n,\mathbf{k}}$ (measured relative to the Fermi surface) in the $n$th layer, $V_{n,n}$ is the Cooper-channel interaction in the $n$th layer and $V_{n,n+1}=V_{n+1,n}^*$ is the amplitude for a Cooper pair to travel from the $n$th to the $(n+1)$th layer. We further use the open boundary conditions: $V_{0,1}=V_{N,N+1}=0$. 
Note that this model does not take into account Coulomb interlayer interactions. This approximation models scenarios in which the interlayer Coulomb interactions are sufficiently screened out. This can happen either as a consequence of the properties of the insulating barriers between the superconducting layers (which can be fine-tuned in systems of artificial stacks of Josephson junctions \cite{klemm_layered_2011}), or near the critical temperature by the presence of a large number of normal electrons \cite{carlson_superconducting_1973}. As one would expect $\omega_J=0$ for neutral superfluids, our approximation should yield accurate results for $\omega_J \ll 2\Delta$. We will heuristically discuss the impact of this contribution where differences in real materials are to be expected. Still, we note that the model adopted here can serve as a pedagogical model that outlines the mechanism of the propagation of collective pseudospin excitations in layered superconductors.

The system is driven by an incident beam of light, represented by the electromagnetic vector potential $\mathbf{A}_n(t)$ polarized in a direction parallel to the layers, which is included via the Peierls substitution in the Hamiltonian \eqref{eq: Hamiltonian}. We adopt the temporal gauge with zero scalar potential. The index $n$ is used to model a layer dependent electromagnetic field, while keeping a uniform electromagnetic field within each layer. 

We proceed by defining the order parameters in the following way:
\begin{eqnarray}
    \Delta_n = \Delta_n^{\mathrm{single}} + \frac{V_{n,n+1}}{V_{n+1,n+1}}\Delta_{n+1}^{\mathrm{single}}+ \frac{V_{n,n-1}}{V_{n-1,n-1}}\Delta_{n-1}^{\mathrm{single}}, \label{eq: order parameters}
\end{eqnarray}
where the single-layer order parameters $\Delta_n^{\mathrm{single}}$ are defined as
\begin{eqnarray}
    \Delta_n^{\mathrm{single}} = -V_{n,n}\sum_{\mathbf{k}}\langle c^\dagger_{\mathrm{k}\uparrow,n}c^\dagger_{-\mathrm{k}\downarrow,n}\rangle. \label{eq: single layer op}
\end{eqnarray}
Note that $\Delta_n$ is the order parameter that defines the energy of quasiparticle excitations and collective modes, whereas $\Delta^{\mathrm{single}}_n$ is simply introduced as a shorthand for the momentum sum over the pair correlation functions of a single layer. After switching to the BCS mean-field description, $\mathcal{H}_N$, becomes (up to a constant)
\begin{eqnarray}
    \mathcal{H}_N &&=\sum_{\mathbf{k}, \sigma, n}\epsilon_{n,\mathbf{k}-e\mathbf{A}_n(t)}c^\dagger_{\mathbf{k}\sigma ,n}c_{\mathbf{k}\sigma, n} \nonumber\\
    &&-\sum_{\mathbf{k},n}\left[\Delta_n^*c^\dagger_{\mathbf{k}\uparrow, n}c^\dagger_{-\mathbf{k}\downarrow, n} + \Delta_n c_{-\mathbf{k}\downarrow,n}c_{\mathbf{k}\uparrow, n}\right]. \nonumber
\end{eqnarray}
We adopt the Anderson pseudospin description \cite{anderson_random-phase_1958}, by defining:
\begin{eqnarray}
     \boldsymbol{\sigma}_{{\mathbf k},n} = \frac{1}{2}\Psi^\dagger_{\mathbf{k},n}\boldsymbol{\tau}\Psi_{\mathbf{k},n} ,
\end{eqnarray}
with $\Psi_{\mathbf{k}, n} = (c_{\mathbf{k} \uparrow ,n},c^\dagger_{-\mathbf{k} \downarrow ,n})^T$ defining the Nambu-spinor of the $n$th layer and $\boldsymbol{\tau}$ being the vector of Pauli matrices. Thus, the Hamiltonian \eqref{eq: Hamiltonian} finally takes the form (up to a constant):
\begin{eqnarray}
    \mathcal{H}_N = \sum_n \mathcal{H}_n^{\mathrm{BdG}}, \quad \mathcal{H}_n^{\mathrm{BdG}} = 2 \sum_{\mathbf{k}}\mathbf{b}_{\mathbf{k},n}\cdot\boldsymbol{\sigma}_{\mathbf{k},n}.
\end{eqnarray}
This is easily recognized as the Bogoliubov-de Gennes (BdG) Hamiltonian of $N$ seemingly-uncoupled superconductors subject to the pseudomagnetic field:
\begin{equation}
    \mathbf{b}_{\mathbf{k}, n} = \left(-\Delta_n', -\Delta_n'', \frac{\epsilon_{n,\mathbf{k}-e\mathbf{A}_n(t)}+\epsilon_{n,\mathbf{k}+e\mathbf{A}_n(t)}}{2}\right). \label{eq: pseudomagnetic field}
\end{equation}
Here, $\Delta_n'$ and $\Delta_n''$ are the real and imaginary parts of the order parameters, respectively. From the Heisenberg equation $\frac{\partial \boldsymbol{\sigma}_{\mathbf{k},n}}{\partial t} = i\left[\mathcal{H}_N, \boldsymbol{\sigma}_{\mathbf{k},n}\right]$, one finds a Bloch-type equation of motion for the expectation values of the pseudospins:
\begin{eqnarray}
    \frac{\partial \langle\boldsymbol{\sigma}_{\mathbf{k},n}\rangle}{\partial t} = 2 \boldsymbol{b}_{\mathbf{k},n} \times \langle\boldsymbol{\sigma}_{\mathbf{k},n}\rangle. \label{eq: BdG-equation}
\end{eqnarray}
All the information about the coupling between individual layers is encoded in the self-consistency condition:
\begin{equation}
    \begin{split}
        \Delta_n' &= -\sum_{\mathbf{k},i_n}V_{n,i_n}\langle\sigma^x_{\mathbf{k}, i_n}\rangle,\\
        \Delta_n'' &= -\sum_{\mathbf{k},i_n}V_{n,i_n}\langle\sigma^y_{\mathbf{k}, i_n}\rangle,
    \end{split} 
    \label{eq: self-consistency}
\end{equation}
where $i_n\in \{n-1,n,n+1\}$ labels the layer itself as well as adjacent layers.

\subsection{Thermal Equiblibrium State}
\begin{figure}
\includegraphics[width=\columnwidth]{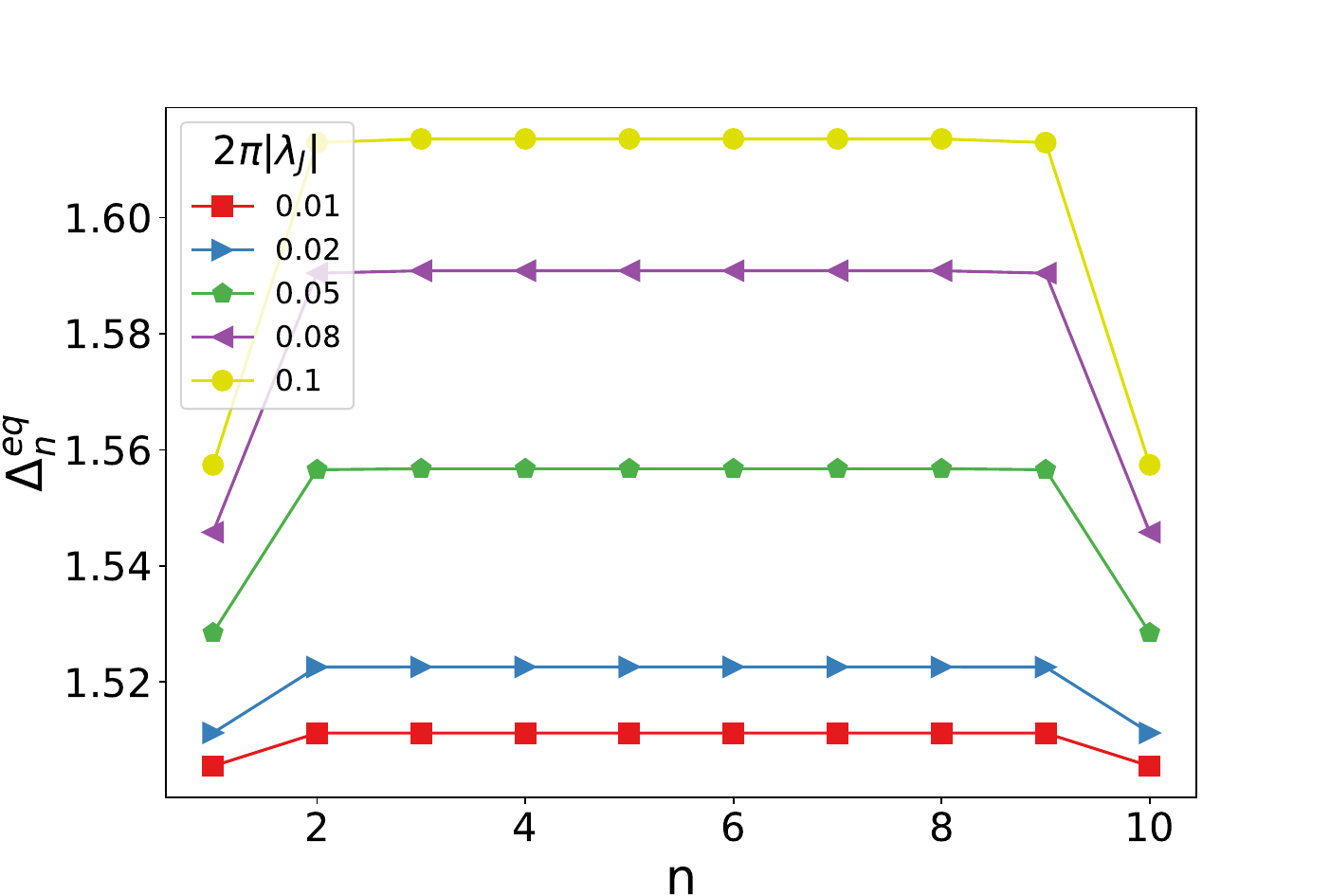}
\caption{Spatial dependence of the superconducting order parameter along the c-axis at $T=0$ with several values of the interlayer tunneling strength $|\lambda_J|$. Here,  $n$ represents the layer index. The dispersion is taken to be $\epsilon_k = -2t\cos(k) +\epsilon_0$ and the parameters are $t=1, U=4, N=10$ and  $\epsilon_0 = 0.8$. The layers consist of one-dimensional chains.}
\label{fig: order parameter gradient}
\end{figure}

In the absence of light, i.e., $\boldsymbol{A}_n=0$ for all $n$, the thermal equilibrium state at temperature $T$ can be found to be \cite{murotani_theory_2017}
\begin{equation}
\begin{split}
     \langle \sigma^{x,\mathrm{eq}}_{\mathbf{k},n}\rangle &= \frac{\Delta^{\mathrm{eq}}_n}{2E_{\mathbf{k},n}}\tanh\left(\frac{E_{\mathbf{k},n}}{2k_BT}\right),\\
    \langle \sigma^{y,\mathrm{eq}}_{\mathbf{k},n}\rangle &= 0,\\
    \langle \sigma^{z,\mathrm{eq}}_{\mathbf{k},n}\rangle &= -\frac{\epsilon_{n,\mathbf{k}}}{2E_{\mathbf{k},n}}\tanh\left(\frac{E_{\mathbf{k},n}}{2k_BT}\right) .
    \label{eq: self-consistency equilibrium}
\end{split}
\end{equation}
Here, we have made use of the $U(1)$ symmetry to choose real and non-negative order parameters. The gap energy $E_{\mathbf{k},n}$ is given by:
\begin{eqnarray}
    E_{\mathbf{k},n} = \sqrt{\epsilon_{n,\mathbf{k}}^{2} + (\Delta^{\mathrm{eq}}_n)^2}.
\end{eqnarray}
In the following, we will focus on symmetrical Josephson junctions with $V_{n,n}=-U$, $V_{n,n+1}=-J, \epsilon_{n,\mathbf{k}}=\epsilon_{\mathbf{k}}$ and $U,J>0$. The interaction strength is then characterized by dimensionless numbers $\lambda_U = N(0)V_{n,n}$ and $\lambda_J = N(0)V_{n,n+1}$, where $N(0)$ is the density of states at the Fermi surface for each layer.
This leads to parallel alignment of the order parameters in thermal equilibrium. Observe that even in equilibrium, the order parameter will exhibit a gradient, since the boundary layers $n=1,N$ only couple to one partner, while the bulk layers ($n\neq 1,N$) exchange Cooper pairs with two neighbors. By numerically solving the equilibrium self-consistency equation \eqref{eq: self-consistency equilibrium} for small enough $V_{n,n+1}$, we find that this dependence can be modelled by (see Fig. \ref{fig: order parameter gradient}):
\begin{align}
    \Delta_n^{\mathrm{eq}} = \begin{cases}
    \Delta_\partial^{\mathrm{eq}} \quad \mathrm{for} \quad n=1,N\\
    \Delta^{\mathrm{eq}} \quad \mathrm{for} \quad n\neq1,N
    \end{cases} ,
    \label{eq: order parameter gradient}
\end{align}
with $\Delta_\partial \neq \Delta$. The pseudospins and single-layer order parameter will exhibit the same inhomogeneity. Equilibrium quantities dressed with the subscript $\partial$ will denote boundary quantities, for which the boundary order parameter $\Delta_\partial$ is to be used. If no such label is present, quantities are to be expressed in terms of the bulk order parameter $\Delta$.
Since the inhomogeneity of the order parameter is restricted to the outermost layers, its influence on the system may be neglected in the limit $N \to \infty$. During the discussion of the collective mode excitations, we will devote special attention to effects related to this inhomogeneity.
\begin{table*}[tp]
\begin{ruledtabular}
\begin{tabular}{ccccc}
  $F(\omega)=\frac{1}{N(0)}\sum_\mathbf{k}\frac{\sigma^x_{\mathbf{k}}}{4E_\mathbf{k}^2-\omega^2}$ && $Y(\omega)=\frac{1}{N(0)}\sum_\mathbf{k}\frac{\sigma^x_{\mathbf{k}}}{4E_\mathbf{k}^2-\omega^2}(\hat{\mathbf{p} \cdot}\nabla_\mathbf{k})^2\epsilon_\mathbf{k}$ &&
  $\Gamma(\omega) = \frac{1}{\Delta}\left(\frac{\Delta^{\mathrm{single}}}{U} + \omega^2 N(0)F(\omega)\right)$\\
  $G(\omega)=(4\Delta^2-\omega^2)N(0)F(\omega)$
  &&$X(\omega)=\sum_\mathbf{k}\frac{4\sigma_\mathbf{k}^x\epsilon_\mathbf{k}}{4E^2_\mathbf{k}-\omega^2}(\hat{\mathbf{p} \cdot}\nabla_\mathbf{k})^2\epsilon_\mathbf{k}$
    &&$\zeta(\omega) = \frac{1}{\Delta}\left(\frac{\Delta^{\mathrm{single}}}{U}-G(\omega)\right)$
\end{tabular}
\end{ruledtabular}
\caption{\label{tab: table1}Definition of functions used throughout this paper. $N(0)$ denotes the density of states at the Fermi surface for each layer and $\Delta^{\mathrm{single}}$ is the single-layer order parameter introduced in Eq. \eqref{eq: single layer op}. These functions are defined in terms of the bulk order parameters/pseudospins. If they appear with a subscript $\partial$, the definition is to be adjusted by performing the substitutions $\Delta\to\Delta_\partial$ and $ \sigma^x_{\mathbf{k}}\to\sigma^x_{\partial,\mathbf{k}}$.}
\end{table*}

\subsection{Optically Driven System}
So far, we have considered the equilibrium state, defined by $\mathbf{A}_n(t)=0$ for all $n$. To describe the interaction with light, we now allow for a non-zero ac-field and derive linearized equations of motion for the real and imaginary parts of the order parameter. In particular, we consider monochromatic light with frequency $\Omega$, where $\mathbf{A}(t)$ is given by: $\mathbf{A}(t)=A(t)\mathbf{\hat{p}}$ with $A(t)=\frac{E}{\Omega}\cdot \sin(\Omega t)$. Here, $\mathbf{\hat{p}}$ is a polarization unit vector in the ab-plane and $E$ is the electric field amplitude.

The interaction with light will drive the system out of equilibrium. For small amplitudes of $\mathbf{A}(t)$, we may write $\Delta_n(t) = \Delta_n^{\mathrm{eq}} + \delta\Delta_n(t)$ where $\frac{\delta\Delta_n(t)}{\Delta_n^{\mathrm{eq}}} \ll 1$ and likewise $\langle\boldsymbol{\sigma}_{\mathbf{k},n}\rangle(t)=\langle\boldsymbol{\sigma}_n^{\mathrm{\mathrm{eq}}}\rangle + \delta\boldsymbol{\sigma}_{\mathbf{k},n}(t)$. The equation of motion \eqref{eq: BdG-equation} is then linearized in the deviations from the equilibrium solution. After a Fourier transformation $t\to\omega$, it takes the following form:
\begin{eqnarray}
    \begin{pmatrix}
    \delta\sigma_{\mathbf{k},n}^x\\
    \delta\sigma_{\mathbf{k},n}^y\\
    \delta\sigma_{\mathbf{k},n}^z
    \end{pmatrix}
    =\boldsymbol{\Sigma}_{\mathbf{k},n}
    \begin{pmatrix}
    -\delta\Delta'_n\\
    -\delta\Delta^{''}_n\\
    \delta b_{\mathbf{k},n}^z
    \end{pmatrix} , \label{eq: linearized BdG}
\end{eqnarray}
    where
\small
    \begin{eqnarray}
    \boldsymbol{\Sigma}_{\mathbf{k},n} =
     - \frac{\sigma_{\mathbf{k},n}^x}{\Delta_n^{\mathrm{eq}}(4E^2_{\mathbf{k},n} - \omega^2)} 
    \begin{pmatrix}
    4\epsilon_{\mathbf{k},n}^2 && 2i\omega \epsilon_{\mathbf{k},n} && 4\Delta_{n}^{\mathrm{eq}}\epsilon_{\mathbf{k},n}\\
    -2i\omega \epsilon_{\mathbf{k},n} && 4E_{\mathbf{k},n}^2 && -2i\omega\Delta_n^{\mathrm{eq}}\\    4\Delta_n^{\mathrm{eq}}\epsilon_{\mathbf{k},n} && 2i\omega\Delta_n^{\mathrm{eq}} &&4 (\Delta_n^{\mathrm{eq}})^2\\
    \end{pmatrix} .\nonumber
\end{eqnarray}
\normalsize
For reasons of readability, we have suppressed the frequency dependence of $\delta\Delta_n, \delta\boldsymbol{\sigma}_{\mathbf{k},n}$ and $\delta b^z_{\mathbf{k},n}$ as well as the label eq and the brackets denoting expectation values, i.e. $\langle\sigma^{x,\mathrm{eq}}_{\mathbf{k},n}\rangle = \sigma^x_{\mathbf{k},n}$

\section{\label{sec:4}Collective Mode Excitations}
In this section, we seek solutions for Eq. \eqref{eq: linearized BdG} near the Fermi surface $\epsilon_\mathbf{k}=0$. In this regime, the real and imaginary parts of the order parameter decouple from each other. Furthermore, we assume the density of states to be constant near the Fermi surface. Since \eqref{eq: self-consistency equilibrium} holds in equilibrium, it is clear that a similar equation must hold for $\delta\Delta_n$ and $\delta\boldsymbol{\sigma}_{\mathbf{k},n}$ to the first order in $\delta$. Substituting \eqref{eq: linearized BdG} into this equation, we find linearized equations that determine $\delta\Delta_n$ self consistently. These solutions will turn out to describe the excitation of amplitude and phase-difference modes.
\subsection{Analytical Description}
If the individual layers are inversion symmetric, we can expand $\delta b^z_{\mathbf{k},n}$  to the lowest order in $A$: $\delta b^z_{\mathbf{k},n} \simeq \frac{e^2}{2}A_n^2(\omega)(\mathbf{\hat{p}}\cdot\nabla_\mathbf{k})^2\epsilon_{\mathbf{k},n}$. Here, $A^2(\omega)$ is the Fourier transform of the squared vector potential.
As discussed above, the dynamics of the real and imaginary part of the order parameter can be approximated by solving a linearized problem in the frequency domain. It turns out that due to the specific structure of the matrix, an analytic solution is available for an arbitrary number of layers $N$.
\subsubsection{Amplitude Oscillations}
For small deviations from equilibrium, we can approximate $|\Delta_n(t)| \approx \Delta_n'(t)$, since the equilibrium order parameter was initialized to be real.
At the Fermi surface, Eq. \eqref{eq: linearized BdG} can be written as a matrix equation:
\begin{align}
T_{ij}\delta\Delta_j(\omega) = -\frac{e^2}{2\zeta(\omega)}\left(\delta_{ii_n}\frac{U}{J}+1\right)A_{i_n}^2(\omega)X_{i_n}(\omega) , \label{eq: e.o.m real part}
\end{align}
where summation over repeated indices is implied, $i_n$ once again labels layers adjacent to the $i$-th layer and and the definitions of $X_{i_n}(\omega)$ and $\zeta(\omega)$ can be found in Table \ref{tab: table1} alongside other functions used throughout this paper. The non-zero elements of the matrix $\mathbf{T}(\omega)$ are given by
\begin{align}
    &T_{ii} = \frac{U\zeta(\omega)-1}{J\zeta(\omega)} \quad i\neq 1,N,\nonumber\\
    &T_{11}=T_{NN}=  \frac{U\zeta_\partial(\omega)-1}{J\zeta(\omega)},\nonumber\\
    &T_{ii+1}=T_{jj-1} = 1 \quad 1\leq i \leq N-2 \quad 2\leq i \leq N,\nonumber\\
    &T_{21}= T_{N-1N} = \frac{\zeta_\partial(\omega)}{\zeta(\omega)}. \nonumber 
\end{align}
\begin{figure}[b]
\includegraphics[width=\columnwidth]{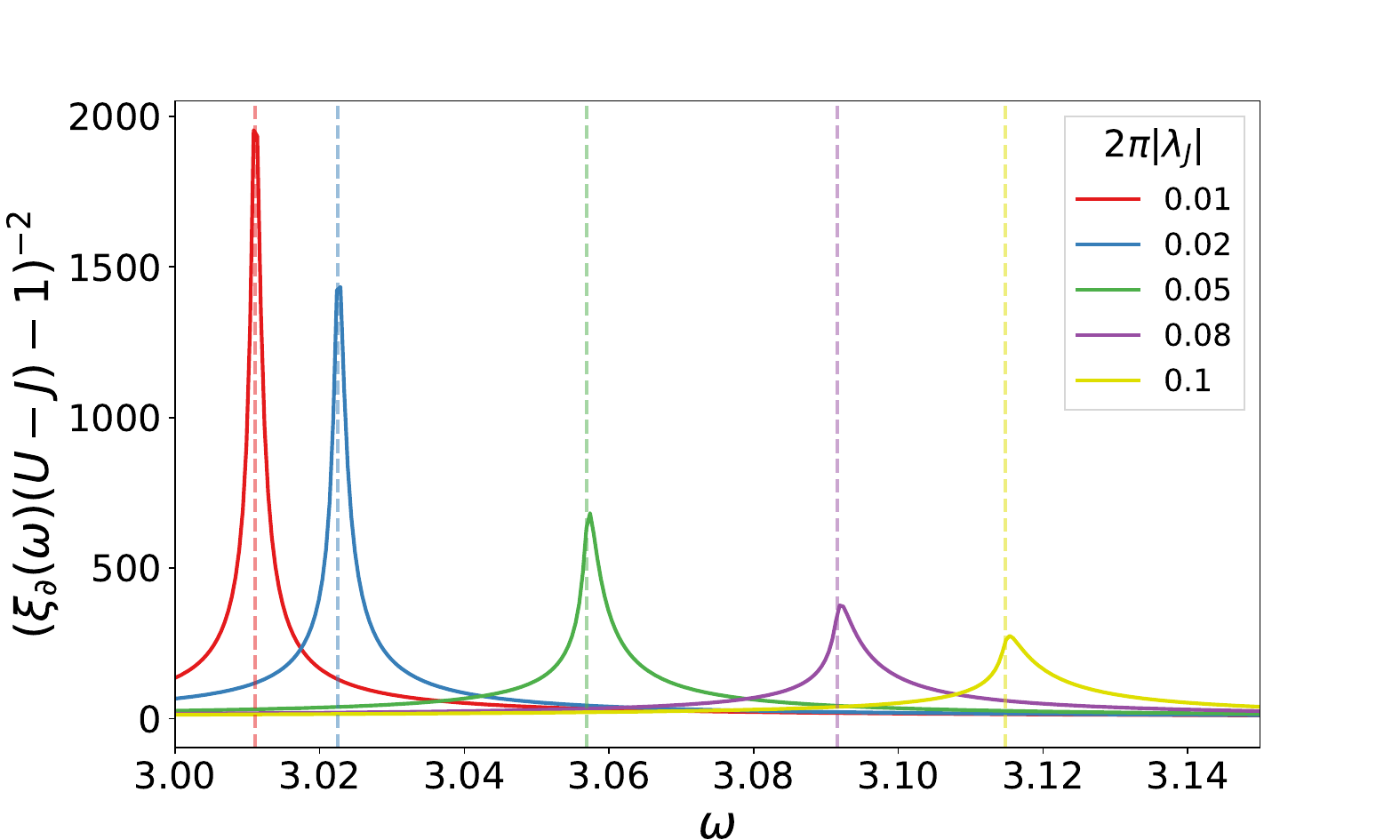}
\caption{\label{fig: boundary Higgs} The function $\left(\zeta_\partial(\omega)(U-J)-1\right)^{-2}$ for several values $|\lambda_J|$.The parameters are the same as in Fig. \ref{fig: order parameter gradient}. The dashed lines indicate values of $2\Delta_\partial$ for the corresponding parameters.}
\end{figure}
The solution to the problem is now readily available by multiplying with the inverse matrix elements. As derived in Appendix \ref{Appendix A}, an analytic expression for these is available for arbitrary $N$. 
To gain a better understanding of the physics of this solutions, we set $A_1(\omega)=A(\omega)$ and $A_n(\omega)=0$ for $2\leq n\leq N$, i.e., we assume that the electromagnetic vector potential only penetrates the first layer. This assumption can be realized in artificial stacks of Josephson junctions, where the thickness of a single layer is typically of the order of the London penetration depth \cite{klemm_layered_2011}, for example in NbN/Al junctions where the thickness of the NbN layers exceeds 390nm \cite{montemurro_enhanced_2023}. In single crystals with intrinsic Josephson junctions like high-$T_c$ cuprates, the layers are much thinner. However, one might still use this approximation to qualitatively describe the expulsion of electromagnetic fields within a limited amount of layers due to the associated computational cost. It is however far from clear, if the physics presented here will carry over to cuprates: First, cuprates are strongly correlated materials. Hence there should be significant deviations from the mean-field limit studied here. Second, the
condition $\omega_J \ll \Delta$ we imposed for the validity of our model is not fulfilled in cuprates
in all the directions, as the gap vanishes along the nodal direction.

In our case, we then arrive at the following solution:
\begin{equation}
\begin{split}
    \delta\Delta_n' &= -\frac{e^2}{2\zeta(\omega)}A^2(\omega)X_\partial(\omega)\left(\frac{U}{J}T^{-1}_{n1} + T^{-1}_{n2}\right). \label{eq: sol real}
\end{split}
\end{equation}
Note that if we apply monochromatic light with frequency $\Omega$ to the first layer, i.e. $A(t) = A\sin(\Omega t)$, with $A=\frac{E}{\Omega}$ and $E$ being the electric field amplitude, then $A^2(\omega)$ is given by
\begin{equation}
    A^2(\omega) = A^2\delta(\omega - 2\Omega),
\end{equation}
leading to oscillations of $\delta\Delta_n'$ with twice the frequency of the applied light. This corresponds to a collective precession of the Anderson pseudospins in the oscillating pseudomagnetic field \cite{tsuji_theory_2015}. Note, however, that even though only the first layer is subject to the incident light, Anderson pseudospin precession persists throughout all layers. The propagation of these oscillations can be explained as follows: The precession of the pseudospins $\boldsymbol{\sigma}_{\mathbf{k},n}$ of the \textit{n}th layer results in oscillations of the order parameter $\Delta_n(t)$, with $\omega = 2\Omega$. The pseudospins create an oscillating pseudomagnetic field, and from Eqs. \eqref{eq: order parameters} and \eqref{eq: pseudomagnetic field} it is clear, that this field also acts on adjacent layers, thereby inducing order parameter oscillations throughout the entire layered structure. 

When ignoring boundary effects (i.e., dropping the subscript $\partial$) the matrix $T_{ij}$ becomes tridiagonal. Consider now $\omega < \omega_+$, where $\omega_{\pm}$ satisfies $ \zeta(\omega_{\pm}) = \frac{1}{U\pm 2J}$ (see Appendix \ref{Appendix A}). At $T=0$, this approximately corresponds to requiring $\omega < 2\Delta^{\mathrm{eq}}$, and thus corresponds to a regime where quasiparticles cannot be excited in the linear regime. One can apply Theorem 2.9 in Ref. \cite{meurant_review_1992} to estimate the decay of the amplitude of the corresponding order parameter oscillations:
\begin{align}
    \frac{\delta\Delta_{n}'}{\delta\Delta_{1}'} \simeq \frac{T_{n 1}^{-1}}{T_{11}^{-1}} \leq e^{-(n-1)\kappa} \label{eq: damping},
\end{align}
where we assumed $\frac{J}{U}\ll1$, showing that the decay is bounded by an exponential function with an exponent $\kappa$ to leading order in $\frac{J}{U}$. The decay constant $\kappa$ is defined by $\frac{U\zeta(\omega)-1}{J\zeta(\omega)} = -2\cosh(\kappa)$.

\subsubsection{Resonance with Higgs mode}
The poles of the inverse matrix elements $T^{-1}_{ij}$ signify the energy of collective amplitude modes. 
These can occur only for $\omega \leq \omega_+$ as otherwise the energy from the drive gets consumed by quasiparticle excitations. It turns out that correspondingly, poles are only found in the regime $\omega<\omega_{+}$ (see Appendix \ref{Appendix A}). For $N \gg 1$, the denominator of $T^{-1}_{ij}$ can be factorized, leading to the following simplified equation for the poles:
\begin{equation}
    \left(\zeta_\partial(\omega)(U-J)-1\right)^2\sinh(N\kappa)\sinh(\kappa)=0 \label{eq: pole-equation-real}
\end{equation}
We find one pole at $\omega=2\Delta^{eq}$ coming from the $\sinh$ factor, which is attributed to the Higgs mode of the bulk superconducting layers. Note that even though the factor $(\left(\zeta_\partial(\omega)(U-J)-1\right))^{-2}$ remains finite, it shows a peak as $\omega \to 2\Delta_\partial^{eq}$ (see Fig. \ref{fig: boundary Higgs}), which is interpreted as the existence of a second Higgs mode associated with the boundary layers.
Hence at $2\Omega = 2\Delta^{eq}$ or $2\Omega = 2\Delta_\partial^{eq}$, the poles of the bulk or boundary Higgs mode and pseudospin precession merge in the expression \eqref{eq: sol real}, leading to the well-known resonant excitation of the Higgs amplitude mode, characterised by a diverging amplitude of the $2\Omega$ oscillations of $\delta\Delta'_n$ \cite{tsuji_theory_2015}. Remarkably, as both the Anderson pseudospin precession and the Higgs mode occur in all the layers, the resonant behavior is not limited to the layer that directly interacts with light. It is worth pointing out that the divergences at $2\Delta^{eq}$ and $2\Delta_\partial^{eq}$ in the bulk layers have different origins. The former is a resonant excitation of the bulk Higgs mode, while the latter originates from the enhanced oscillation amplitude of the pseudomagnetic field from resonance in the boundary layer.
Furthermore we can observe that the factor $X_\partial(\omega)$ diverges for $2\Delta_\partial^{eq} \leq \omega \leq \max_{\mathbf{k}}\{2E_{\partial,\mathbf{k}}\}$. 
This is interpreted as the incoherent oscillations from quasiparticles in the top layer $\Delta_\partial^{eq}$ inducing oscillations in lower layers by the same mechanism inducing pseudospin precession in the lower layers.

\subsubsection{Phase Oscillations}
Similarly to the amplitude of the order parameter, for small deviations from equilibrium, we have the phase $\phi_n(t) = \mathrm{arg}(\Delta_n(t))$ proportional to the imaginary part of the order parameter: $\delta\phi_n \simeq \frac{\delta\Delta_n''}{\Delta_n}$.
This time, we restrict ourselves to frequencies $\omega\leq\omega_{\mathrm{max}}$,
where $\omega_{\mathrm{max}}$ is the solution of $
\Gamma(\omega_{\mathrm{max}})=\frac{1}{U-2J}$ and for the definition of $\Gamma(\omega)$ we refer to Table \ref{tab: table1}. This is the regime where one should expect collective phase-difference modes. To see this, one can approximate $\omega_{\mathrm{max}}\approx \omega_L = 8(\Delta^{\mathrm{eq}})^2\frac{4\lambda_J}{\lambda_U^2-4\lambda_J^2}$, which corresponds to the energy of the Leggett mode in two-band superconductors with the interband coupling $2\lambda_J$ \cite{leggett_number-phase_1966}. This energy corresponds to that of an oscillating relative phase between every pair of adjacent layers. Lower momentum phase-difference modes will involve some adjacent order parameters oscillating in phase, thereby reducing the energy. For a derivation of this condition, see Appendix \ref{Appendix A}. 

We adopt the convention, where the interlayer phase mode is referred to as 
\textit{phason} in the system without inter-layer Coulomb interactions that we are studying in this paper, as opposed to the Josephson plasma mode in the presence of long-range inter-layer Coulomb interactions.
The system of linear equations governing the time evolution of $\delta\Delta_n''$ is then found to be:
\begin{align}
T_{ij}\delta\Delta_j(\omega) = -\frac{\omega e^2}{i\Gamma(\omega)}\left(\delta_{ii_n}\frac{U}{J}+1\right)A_{i_n}^2(\omega)Y_{i_n}(\omega) \label{eq: e.o.m imaginary}
\end{align}
 The non-zero elements of the matrix $T_{ij}(\omega)$ are given by
\begin{align}
    &T_{ii} = \frac{U\Gamma(\omega)-1}{J\Gamma(\omega)} \quad i\neq 1,N,\nonumber\\
    &T_{11}=T_{NN}=  \frac{U\Gamma_\partial(\omega)-1}{J\Gamma(\omega)},\nonumber\\
    &T_{ii+1}=T_{jj-1} = 1 \quad 1\leq i \leq N-2 \quad 2\leq i \leq N,\nonumber\\
    &T_{21}= T_{N-1N} = \frac{\Gamma_\partial(\omega)}{\Gamma(\omega)}. \nonumber 
\end{align}

As in Eq. \eqref{eq: sol real}, we study the solution for $A_1\neq0$:
\begin{align}
    \delta\Delta_n'' = \frac{\omega e^2}{i\Gamma(\omega)}Y_\partial(\omega)A^2(\omega)\left(\frac{U}{J}T^{-1}_{n1} + T^{-1}_{n2}\right). \label{eq: sol Imaginary}
\end{align}
The solution for the phase has a similar structure to the amplitude, featuring tunneling induced Anderson pseudospin precession and light induced pair-hole excitations.
Unlike for the amplitude, the lattice filling has a significant influence on the phase-difference oscillations: We expand $\frac{\partial^2\epsilon_\mathbf{k}}{\partial k^2}$ in powers of $\epsilon$ to obtain:
\begin{align}
   Y_\partial(\omega)&= \int_{-\infty}^{\infty}d\epsilon \frac{\Delta\tanh(\sqrt{\epsilon^2+(\Delta^{\mathrm{eq}}_\partial)^2}/2k_BT)}{2\sqrt{\epsilon^2+(\Delta^{\mathrm{eq}})^2}(4\epsilon^2+4(\Delta_\partial^{\mathrm{eq}})^2-\omega^2)}\nonumber\\
   &\times \left(c_0+c_1\epsilon+c_2\epsilon^2 +...\right).
\end{align}
Note that for a cosine dispersion $\epsilon_\mathbf{k}=\epsilon_0-2t\sum_{k_i}\cos(k_i)$ at half filling ($ \epsilon_0=0$), we have $c_0=0, c_1=-1$ and $c_n=0$ for $n>1$. Since the term linear in $\epsilon$ vanishes due to symmetry, phase oscillations are only obtained for partial filling, which is consistent with previous studies \cite{cea_nonlinear_2016}.
\subsubsection{Resonance with phase-difference modes}
Compared to the amplitude mode, elements of the inverse matrix $T_{ij}^{-1}$ now take a different form (see Appendix \ref{Appendix A}). Hence the eigenspectrum of the phase differs greatly from that of the amplitude.

In the limit $N\to \infty$ one can once again factor the denominator of the inverse matrix elements and then find an equation for the poles:
\begin{equation}
\left(\Gamma_\partial(\omega)(U-J)-1\right)^2\sin(N\kappa)\sin(\kappa)=0 ,
    \label{eq: equation for phase modes}
\end{equation}
where this time we have $-2\cos(\kappa) = \frac{U\Gamma(\omega)-1}{J\Gamma(\omega)}$.

To find solutions to the above equation, we introduce an approximation for the function $F(\omega)$ (see Table \ref{tab: table1}), which also yields an approximation for $\Gamma(\omega)$.
Note that close to the Fermi surface, we can replace the summation over the momenta $\mathbf{k}$ by an integral, $\sum_{\mathbf{k}}\to N(0)\int_{-\omega_D}^{\omega_D}d\epsilon$ where $\omega_D$ is the Debye frequency and $N(0)$ is the density of states at the Fermi surface. In the BCS approximation one can then proceed by replacing the boundaries of the integrals with infinity, i.e., $\int_{-\omega_D}^{\omega_D} \to \int_{-\infty}^{\infty}$. Note also, that since the phase-difference modes are typically of frequencies $\omega\ll2\Delta^{\mathrm{eq}}$, we can approximate at $T=0$ as follows:
\begin{eqnarray}
    F(\omega)&&=\int_{-\infty}^{\infty}d\epsilon \frac{\Delta\tanh(\sqrt{\epsilon^2+(\Delta^{\mathrm{eq}})^2}/2k_BT)}{2\sqrt{\epsilon^2+\Delta^{(\mathrm{eq}})^2}(4\epsilon^2+4(\Delta^{\mathrm{eq}})^2-\omega^2)} \nonumber \\
    &&\approx\frac{1}{4\Delta^{\mathrm{eq}}}. \label{eq: approximation} 
\end{eqnarray}
Using Eq. \eqref{eq: approximation} an analytic solution for Eq. \eqref{eq: equation for phase modes} is available at $T=0$, yielding a discrete excitation spectrum with frequencies given by
\begin{equation}
    \omega^2_m =\frac{8|\lambda_J|(\Delta^{\mathrm{eq}})^2 (1-  \cos(\frac{\pi m}{N}))}{ (2 |\lambda_J| + |\lambda_U|) (|\lambda_U| + 2 |\lambda_J| \cos(\frac{\pi m}{N}))}\label{eq: discrete frequencies},
\end{equation}
where $m = 0,1,\ldots N$. This yields $N+1$ nonzero solutions for $\omega_m$. However, $\omega_N$ is also a root of the numerator of $T_{ij}^{-1}$, and thus there are $N$ collective phase-difference modes $\omega_0, \ldots,\omega_{N-1}$, where $\omega_{0}=0$ corresponds to the Goldstone mode. Also, Eq. \eqref{eq: discrete frequencies} only depends on $\Delta^{\mathrm{eq}}$ and not on $\Delta_\partial^{\mathrm{eq}}$, suggesting that the spectrum is independent of the boundary conditions.\footnote{One might think about ignoring the boundary effects from the beginning. However, if $\Delta_\partial^{\mathrm{eq}} = \Delta^{\mathrm{eq}}$ from the get-go, then one obtains the same energies as \eqref{eq: discrete frequencies} but replacing $\frac{\pi m}{N}\to\frac{\pi m}{N+1}$, which would increase the number of modes from $N$ to $N+1$. This would be unphysical, as there are only $N$ phase degrees of freedom.} 
These excitations can be understood as follows: The physical effect of the Josephson coupling is, to align the order parameters parallel to each other. If the order parameter of a layer is rotated away from the parallel alignment, the attractive potential of the Josephson coupling aims to realign adjacent order parameters, leading to harmonic oscillations. The eigenmodes of these coupled harmonic oscillators are the phason modes.   
We observe that independent of the system size, the energy of the collective phase excitations has an upper bound of:
\begin{equation}
    \omega_{m}^2 < \omega^2_{N} = \omega^2_{\mathrm{max}} = \frac{|\lambda_J| (\Delta^{\mathrm{eq}})^2}{(|\lambda_U|^2-4|\lambda_J|^2)}.
\end{equation}
For finite system sizes, the phason mode remains gapped, with the energy of the gap decreasing with increasing the number of layers, yielding a gapless phason in the interval $(0,\omega_{\mathrm{max}}]$   as $N\to \infty$. 
The pole defined by $\Gamma_\partial(\omega)(U-J)=1$ encapsulates boundary effects. However, it lies within the aforementioned  continuum and thus yields no new excitation. When considering the presence of interlayer Coulomb interactions, a gap is expected to open up in the dispersion of the phason mode, corresponding to the Josephson plasma edge $\omega_J$ \cite{fertig_collective_1991}. As discussed, their omission may be justified if $\omega_J \ll 2\Delta^{\mathrm{eq}}$.
Interlayer Coulomb interactions should be considered for real materials, however, they usually leave the amplitude mode unchanged. Their effect on the phase mode may be included in this formalism by a perturbation to the T-matrix, $T_{ij} \to T_{ij} +T^{(1)}_{ij}$. Applying the first order perturbation theory for linear operators, the eigenvalues change $\omega_m \to \omega_m + \omega^{(1)}_m$, which will open up the gap. Still, Anderson's pseudospin resonance will occur when the poles of the inverse matrix match $2\Omega$. Hence the mechanism of Anderson pseudospin resonance with the phase mode should also be present in the presence of interlayer Coulomb interaction, but with a change in the dispersion relation. This, however, should be investigated by future research, especially because $\omega_J$ is typically of the order of GHz to THz, i.e. of the same order of magnitude as $\Delta^{\mathrm{eq}}$. Hence, the first order perturbation theory might not be sufficient in the cases of interest.
\begin{figure}
\includegraphics[width=\columnwidth]{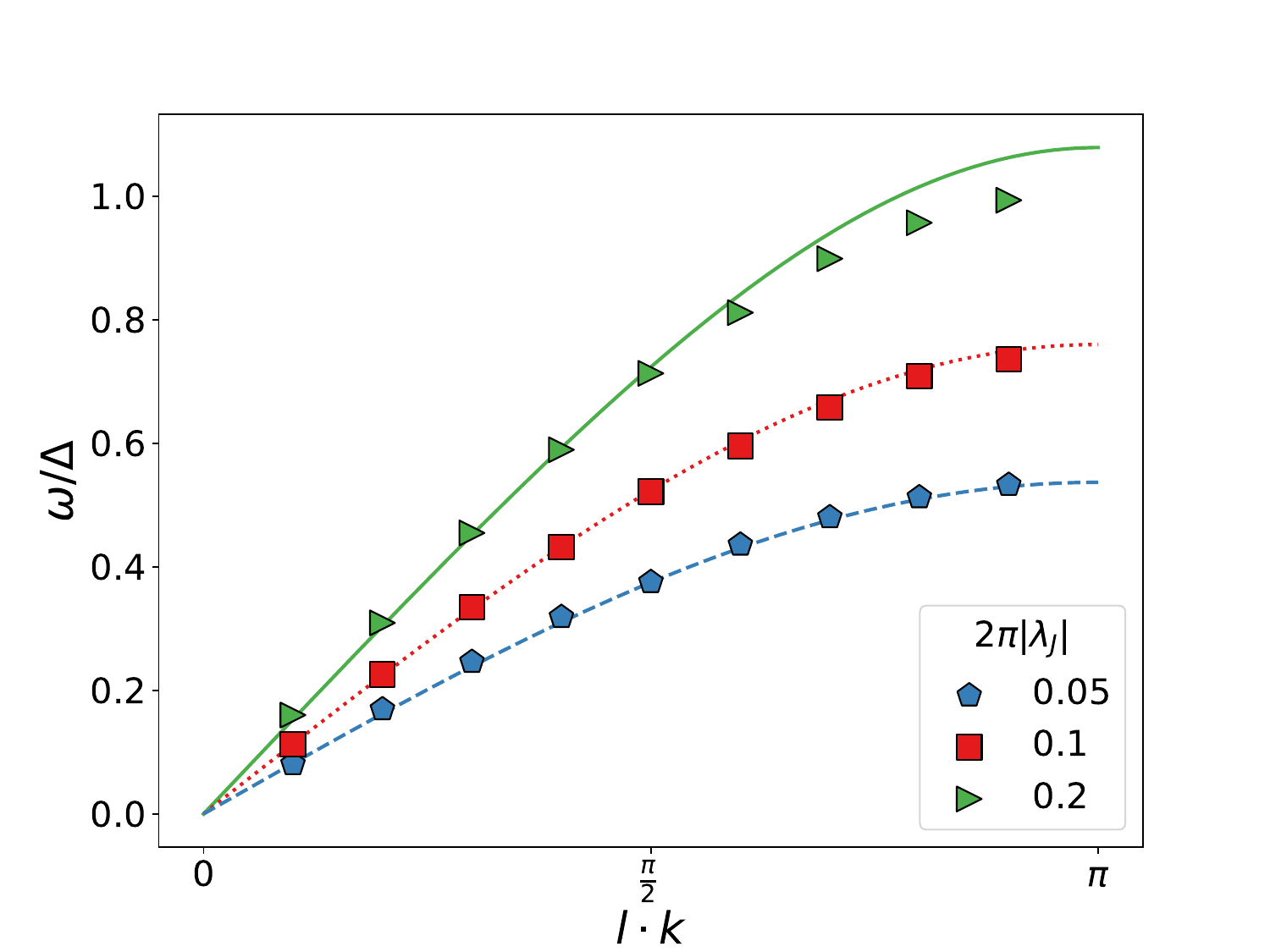}
\caption{\label{fig: dispersion} Dispersion relation of the phason at $T=0$. The dots are numerically obtained. Lines are plots of Eq. \eqref{eq: dispersion_phason} for a system with $N=10$ for various values of the interlayer tunneling strength $|\lambda_J|$. The other parameters are the same as in Fig. \ref{fig: boundary Higgs}.}
\end{figure}

To sharpen our understanding of the phason mode, it is necessary to find the dependence of the eigenenergies \eqref{eq: discrete frequencies} on the momentum $k$ along the c-axis. For this purpose, we consider the normal modes, i.e. the generalized eigenvectors $\delta\Delta^{''\mathrm{norm}}_m \in \mathrm{ker}(\mathbf{T}(\omega_m))$. As we can neglect boundary effects in the large $N$ limit, we assume $\Delta_\partial^{\mathrm{eq}}=\Delta^{\mathrm{eq}}$. In this case, the T-matrix becomes a tridiagonal, symmetric Toeplitz matrix, for which one can show, that the eigenvectors are of the following form \cite{noschese_tridiagonal_2013}:
\begin{equation}
\delta\Delta^{\mathrm{norm}''}_m(x_n,t) \propto \sin(x_n\frac{m\pi}{L})\cos(\omega_m t) \label{eq: normal modes},
\end{equation}
where we assumed the thickness of the entire stack (including superconducting layers and insulating barriers) to be $L$. Hence $x_n = \frac{L}{N+1}n \approx \frac{L}{N}n$  is the position of the $n$th layer measured along the c-axis. 
Eq. \eqref{eq: normal modes} represents standing wave solutions to the wave equation with momentum $k=\frac{m\pi}{L}$ along the c-axis. Here $\omega_m$ refers to the eigenvalues of the problem without boundary effects. However, based on our discussion above, we replace them by the actual ones in order to obtain the correct number of excitations. Furthermore, \eqref{eq: normal modes} confirms the intuition, presented in Sec.~\ref{sec:2}, that the Goldstone mode with $m=0$ corresponds to a global in-phase rotation of all the order parameters.
Hence at $T=0$, we have the following dispersion relation for the phason mode:
\begin{equation}
    \omega^2_{k} =\frac{(8|\lambda_J|(\Delta^{\mathrm{eq}})^2 (1-  \cos(l\cdot k)))}{
 (2 |\lambda_J| + |\lambda_U|) (|\lambda_U| + 2 |\lambda_J| \cos(l\cdot k))} , \label{eq: dispersion_phason}
\end{equation}
where $l=\frac{L}{N}$ is the thickness of the insulating barriers between the superconducting layers.
For small $k$, the dispersion becomes linear:
\begin{equation}
    \omega \simeq c_P k, \quad c_P =\sqrt{|\lambda_J|} \frac{2\Delta^{\mathrm{eq}} l}{2|\lambda_J| + |\lambda_U|} ,
\end{equation}
where $c_P$ is the phason sound velocity. In the presence of the interlayer Coulomb interaction, described by a perturbed T-matrix with matrix elements $T_{ij} + T^{(1)}_{ij}$, the eigenvectors associated with the first order corrections of the eigenvalues are precisely the unperturbed eigenvectors. Hence the normal modes are stable to the first order in the perturbation.
\begin{figure}
    \centering    
    \includegraphics[width=1.0 \columnwidth]{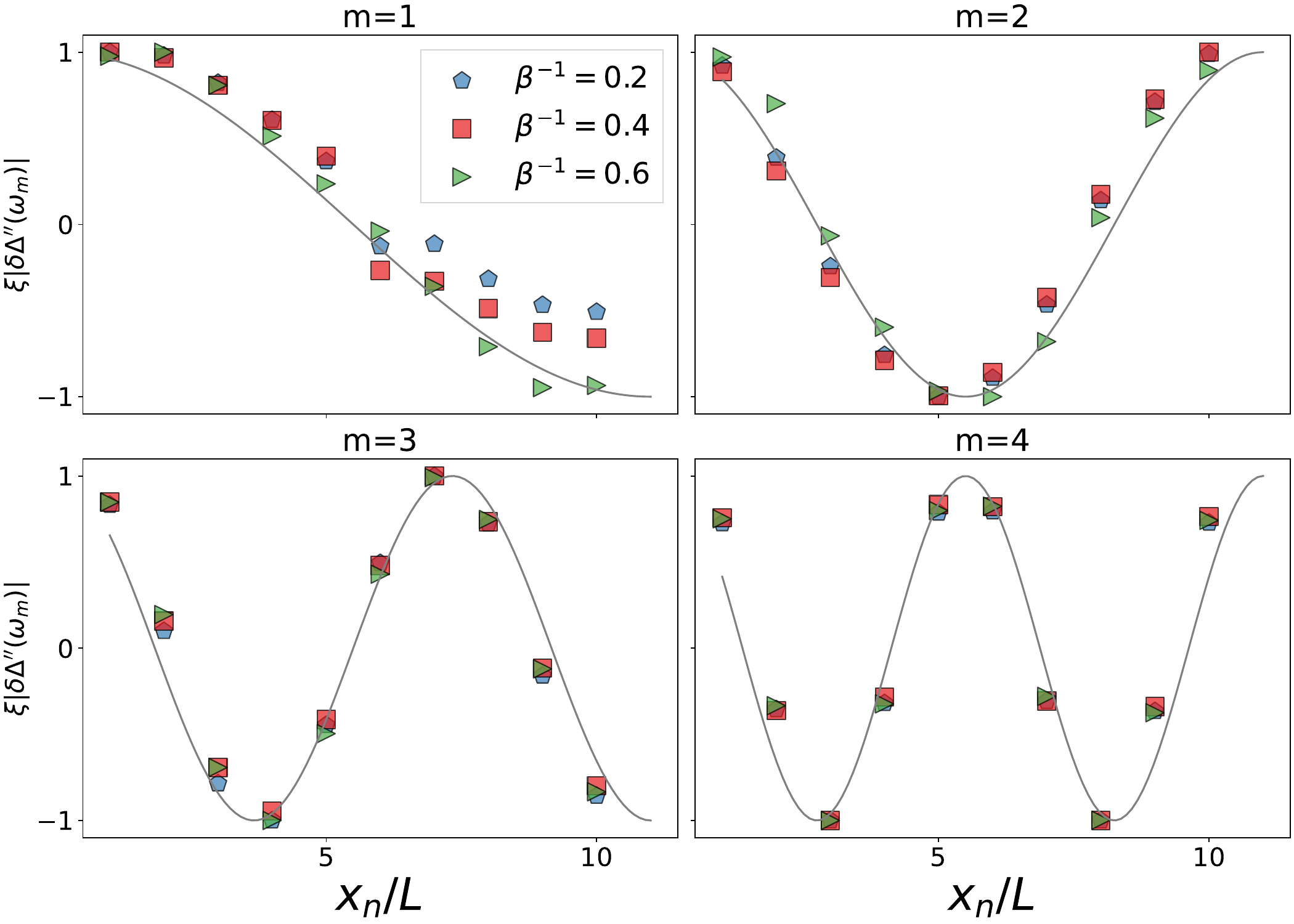}
    \caption{\textbf{(a)}-\textbf{(d)}: The oscillation amplitude $\xi|\delta\Delta''_n(\omega_m)|$ for $m=1,\ldots,4$, where $\xi = \mathrm{sign}(\Re(\delta\Delta''_n(\omega_m)))$ as determined by simulations. The 10-dimensional vectors are normalized by $||\cdot||_\infty$. Grey lines show the normal modes with momentum $k_m = m\frac{\pi}{L}$ from Eq.\eqref{eq: normal modes}.}
    \label{fig: momentum}
\end{figure}
\begin{figure}
    \centering
    \includegraphics[width=\columnwidth]{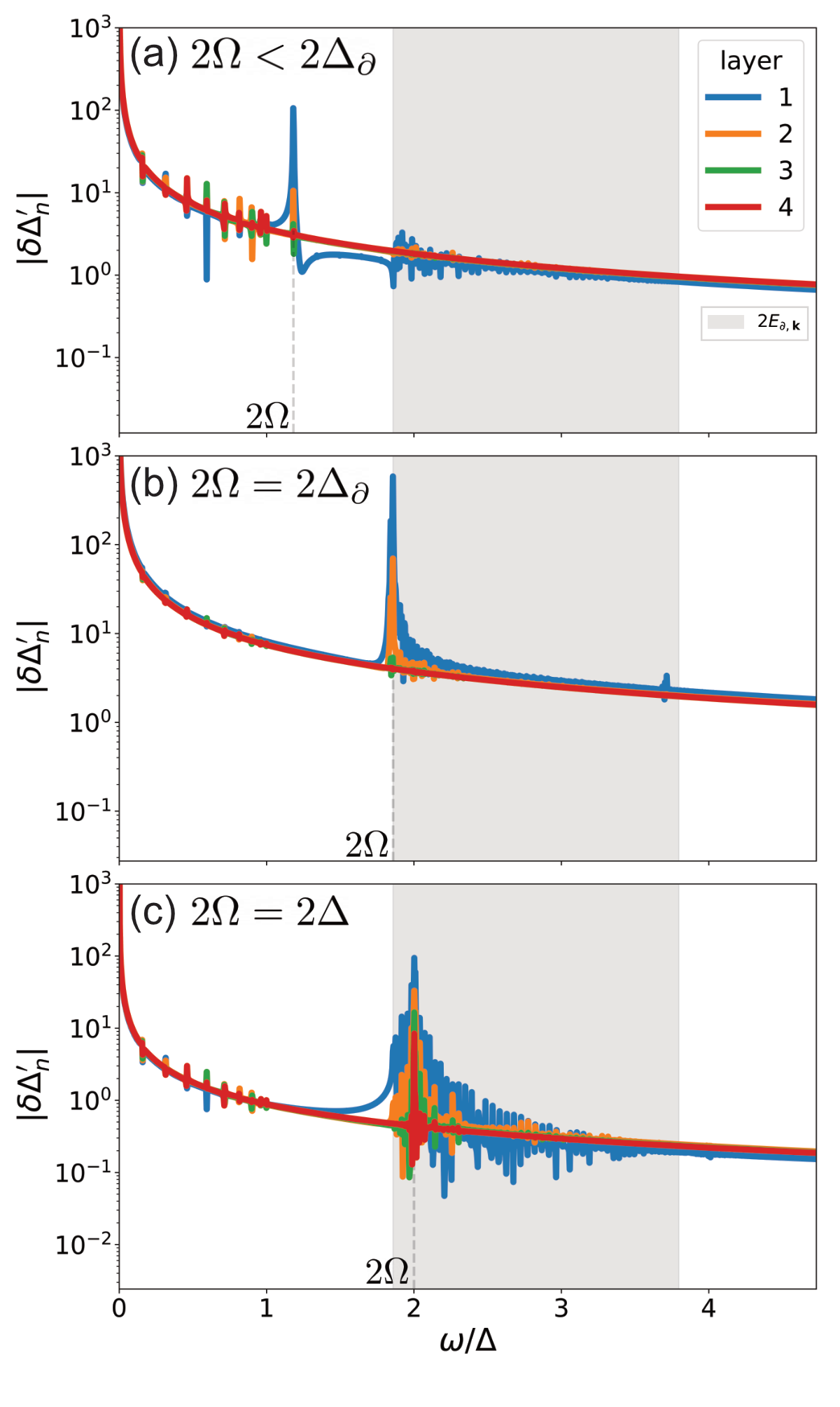}
    \caption{Numerical results from long time simulations ($t_{\mathrm{max}}=1000$) for $\delta\Delta'_n(\omega), \quad 1\leq n \leq 4$ for different values of the driving frequency $\Omega$ where only the first layer was subject to the drive. Parameters are taken to be $\epsilon_0 = 0.8, t=1, U=4, E=0.15, N=10$, and \textbf{(a)} $2\Omega = 1.2\Delta^{\mathrm{eq}}$, \textbf{(b)} $2\Omega = 2\Delta_\partial^{\mathrm{eq}}$, and \textbf{(c)} $2\Omega = 2\Delta^{\mathrm{eq}}$. The grey shaded area marks the quasiparticle spectrum of the boundary layers.}
    \label{fig:real part simulations}
\end{figure}
\subsection{Numerical Results}

Numerical simulations are performed through the self-consistent second-order Runge-Kutta algorithm for the equation of motion \eqref{eq: BdG-equation}. For the sake of computational efficiency, we assume that each layer consists of a one-dimensional chain with the dispersion $\epsilon_k=-2t \cos(k)+\epsilon_0$. This is expected to yield qualitatively similar results as two-dimensional layers for processes near the Fermi-surface, as the dimension only enters via k-space integration, which can be replaced by an energy integral $\sum_\mathbf{k} \to N(0)\int_{-\infty}^{\infty}$ with the constant density of states $N(0)$ near the Fermi surface. Hence, the dimension of the layers enters the formalism only via the density of states at the Fermi surface and can thus be absorbed into the amplitude of the hopping term. The parameters are taken to be $t=1, \epsilon_0 = 0.8, U=4, E=0.15$ and $N=10$. To achieve a good resolution in the frequency space, we conducted long-time simulations ($t_{\mathrm{max}}=1000$). We study a more realistic case of shorter pulses in Section \ref{sec:5} when discussing the third-harmonic generation.  We apply the ac field only to the first layer.

Figure \ref{fig: dispersion} shows the $T=0$ dispersion relation for the phason (Eq. \eqref{eq: discrete frequencies}) for different values of the interlayer coupling compared to the frequencies of the phase modes as determined by the real-time simulation. Note that the simulation and calculation disagree as $\omega$ increases, since here the approximation in Eq. \eqref{eq: approximation} fails.

In Fig. \ref{fig: momentum}, the oscillation amplitude of the order parameter phase in each layer is shown at resonance with the four lowest energy phason modes. The grey lines indicate the analytical result for the corresponding eigenmodes from Eq. \eqref{eq: normal modes}. As Eq. \eqref{eq: normal modes} was obtained by setting $\Delta_\partial^{\mathrm{eq}} = \Delta^{\mathrm{eq}}$, Fig. \ref{fig: momentum} demonstrates that boundary effects do not significantly influence the eigenmodes.\\
\indent Figure \ref{fig:real part simulations} shows representative plots of the Fourier transform $\delta\Delta_n'(\omega)$ of the real part of the order parameters of the first four layers for different frequencies of the periodic drive.
If $2\Omega$ does not match the resonance condition of the Higgs mode (see Fig. \ref{fig:real part simulations}(a)), we observe a dominant peak at $2\Omega$ in the first layer which we attribute to pseudospin precession. In accordance with our analytical predictions, there are visible, though quickly decaying peaks at the same frequencies in the underlying layers, demonstrating numerically the propagation of light-induced pseudospin precession. Additionally, there is a band of peaks starting at $2\Delta_\partial^{\mathrm{eq}}$ which can be attributed to the predicted light-induced quasiparticle excitations. There are also nine smaller peaks at frequencies smaller than $2\Omega$. Their number and frequency matches that of the phason modes. For long time simulations, processes involving particles away from the Fermi surface occur. Here, the real and imaginary parts are coupled in the linearized BdG-equation \eqref{eq: linearized BdG} leading to an interaction of the amplitude and phase modes.\\
\indent At $2\Omega = 2\Delta_\partial^{\mathrm{eq}}$ (Fig. \ref{fig:real part simulations}(b)) one observes a resonance with the Higgs mode of the boundary layer, signified by a strongly enhanced oscillation amplitude. This also results in amplified oscillations in other layers. However, their amplitude still remains damped compared to the amplitude of the first layer. The behavior qualitatively changes at $2\Omega=2\Delta^{\mathrm{eq}}$ (Fig. \ref{fig:real part simulations}(c)), where the resonance condition for the bulk layers is fulfilled. Here, all the bulk layers feature a resonant amplification of the order parameter oscillations, even if they do not directly interact with the incident beam of light.
\section{Josephson Current and Third Harmonic Generation \label{sec:5}}
\begin{figure}[H]
    \centering
    \includegraphics[width=\columnwidth]{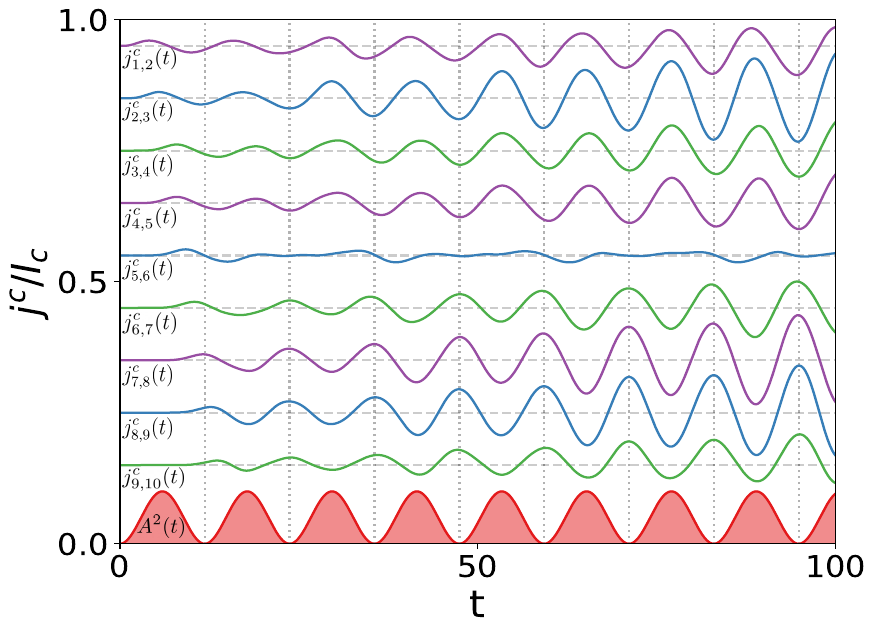}
    \caption{Josephson current $j^c_J/I_c$ between adjacent layers for a ten-layer system at $2\Omega = \omega_1$. Parameters are $J=0.2, \epsilon_0 = 0.8, \beta^{-1}=0.4, U=4, t=1$. All the currents are centered around 0 (grey dashed lines), but have been shifted up and down for visibility. The vector potential is plotted in arbitrary units, and the dotted lines are a guide to the eye, representing the minima of $A^2(t)$.}
    \label{fig: Josephson Current}
\end{figure}
Due to the characteristic anisotropy of layered superconductors, the current response to the interaction with light also exhibits a strong anisotropy. In particular, one can differentiate between the interlayer current along the c-axis, i.e. the Josephson current, and an intralayer current that flows within the ab-plane of each layer. 
The former is given by $j^c_{n,n+1} = I_c\sin(\phi_{n}-\phi_{n+1}) $, where $\phi_n$ is the phase of the order parameter of the \textit{n}th layer and $I_c$ is a parameter of the Josephson junction known as the Josephson critical current. As we noted before, notable oscillations of the phase are only obtained at partial filling. Figure \ref{fig: Josephson Current} shows this current at the point of resonance $2\Omega = \omega_1$ (see Eq. \eqref{eq: discrete frequencies}) for a ten-layer system. After a few cycles of the ac field, in-phase oscillations of the current start to emerge, with frequency $2\Omega$ throughout the entire stack. Due to the resonance, their amplitude is rapidly increasing. We observe that the currents in the bottom half of the stack ($n=6, \cdots, 10$) exhibit a phase shift of $\pi$ with respect to the currents in the top half ($n=1, \cdots, 5$). This is understood by the fact that the $\omega_1$ mode has a node (see Fig. \ref{fig: momentum}(a)) in the middle of the stack, resulting in different signs of the oscillation amplitude in the upper and lower half of the stack. 

Apart from the Josephson current, that flows normal to the layers, there is also a current within the ab-plane. The layer dependent current operator is given by \cite{murotani_theory_2017}
\small
\begin{align}
    \mathbf{j}_n^{\mathrm{ab}}(t)&=e\sum_{\mathbf{k}\sigma}\nabla_{\mathbf{k}}\epsilon_{n,\mathbf{k}-e\mathbf{A}_n(t)}c^\dagger_{\mathbf{k}\sigma, n}c_{\mathbf{k}\sigma, n}\nonumber \\
    &=e\sum_{\mathbf{k}\sigma}\nabla_\mathbf{k}\left(\epsilon_{n,\mathbf{k}-e\mathbf{A}_n(t)}-\epsilon_{n,\mathbf{k}-e\mathbf{A}_n(t)}\right)\left(\sigma^z_{\mathbf{k},n}(t) +\frac{1}{2}\right) \nonumber\\
    &\quad + \mathrm{const.}, 
\end{align}
\normalsize
where the last term is constant in the mean-field approximation.
By writing $\sigma^z_{\mathbf{k},n}(t) = \sigma^{z,\mathrm{eq}}_{\mathbf{k},n} + \delta\sigma^z_{\mathbf{k},n}(t)$ we obtain the light induced nonlinear current:
\begin{align}
    \mathbf{j}^{ab,(3)}_n(t) = -2eA_n(t)\sum_{\mathbf{k}}\nabla_\mathbf{k}\left(\mathbf{\hat{p}}\cdot \nabla_\mathbf{k}\right)\delta\sigma^z_{\mathbf{k},n}(t).
\end{align}
 Note that since the  current is proportional to $ A_n(t)$, only layers that are illuminated by light exhibit this oscillating current, which is in contrast to the Josephson current, that penetrates through the entire stack at resonance with the phason modes. 
Observe that $\delta\sigma^z_{n,\mathbf{k}}(t)$ oscillates with $2\Omega$ due to the precession of Anderson's pseudospins. Thus the current exhibits oscillations with $3\Omega$, and since the emitted radiation is proportional to this current, third-harmonic generation (THG) can be observed.

Since we simulate one-dimensional layers, the polarization vector $\mathbf{\hat{p}}$ is naturally fixed. Using the linearized BdG equation in Fourier space, Eq. \eqref{eq: linearized BdG}, we then obtain the following expression \cite{murotani_theory_2017}:
\small
\begin{align}
    j^{ab,(3)}_n(t) = -2eA_n^2(t)\int_{-\infty}^{\infty} \frac{d\omega}{2\pi}e^{i\omega t}\sum_{\mathbf{k}}\frac{\partial^2 \epsilon_\mathbf{k}}{\partial k^2}\frac{\sigma_{n,\mathbf{k}}^x}{4E_{n,\mathbf{k}}^2-\omega^2} \nonumber\\
    \times\left(4\epsilon_{\mathbf{k}}\delta\Delta_n'(\omega)+2i\omega\delta\Delta''_n(\omega)-2\Delta_n^{\mathrm{eq}}e^2A^2(\omega)\frac{\partial^2 \epsilon_\mathbf{k}}{\partial k^2}\right).
    \label{eq: THG-current}
\end{align}
\normalsize
\begin{figure}
    \centering
    \includegraphics[width=\columnwidth]{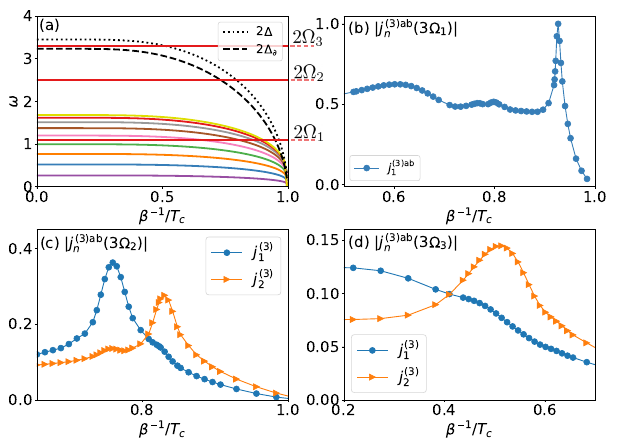}
    \caption{\textbf{(a)} Energy of the phason and Higgs modes compared to $2\Omega$ depending on temperature. Parameters are $U=4, J=0.2, \epsilon_0 = 0.8$, and $t=1$.  \textbf{(b)} Nonlinear current in arbitrary units for $A_1\neq 0, A_n = 0\; (n>1)$, and $\Omega = \Omega_1$. \textbf{(c)}-\textbf{(d)}: Nonlinear current $j^{ab,(3)}_n$ normalized by the largest amplitude for $\Omega = \Omega_{2,3}$ and $A_{1}=A_{2}\neq 0, A_n=0\; (n>2)$. For all the simulations, illumination time was $t_{\mathrm{max}}=50$.}
    \label{fig: THG-current}
\end{figure}
This shows that the THG current splits into three different contributions. These three summands are attributed to: Amplitude mode, phase mode and quasiparticle excitations, respectively. At resonances with the Higgs and phason modes, the amplitude of their oscillations grows, leading to resonantly enhanced THG. This was first experimentally demonstrated for NbN by Matsunaga \textit{et al.} \cite{matsunaga_light-induced_2014, matsunaga_polarization-resolved_2017}. Note that the resonant enhancement of THG at $2\Omega = 2\Delta$ is attributed both to quasiparticle excitations and the Higgs mode, where the contribution of the former depends on the polarization of the incident light~\cite{cea_nonlinear_2016}. However, in the one-dimensional case presented here, the polarization dependence is not present. Even though in the clean system studied here the quasiparticle contribution is dominant \cite{cea_nonlinear_2016, murotani_theory_2017}, in the presence of disorders the Higgs contribution may become dominant \cite{tsuji_higgs-mode_2020, murotani_nonlinear_2019,shimano_higgs_2020}.
The resonance of the THG is mediated by $2\Omega$ oscillation of pseudospins. Such a coherent $2\Omega$ oscillation has been observed in a pump-probe spectroscopy for both NbN \cite{matsunaga_light-induced_2014} and high-$T_c$ cuprates \cite{Katsumi_higgs_2018}.

Figure \ref{fig: THG-current}(a) shows the temperature dependence of the energy of the collective modes (Higgs modes in black and phason modes in color) as well as $2\Omega$ for three different frequencies of the incident light ($\Omega_i, \; i \in \{1,2,3\}$). At temperatures where these lines cross, our theory predicts resonant enhancement of the nonlinear current.

At $\Omega = \Omega_1 = 0.55$, we choose a single layer illumination, i.e. $A_1\neq 0, A_n = 0 \;(\forall n>1)$. We find three peaks associated with the resonant excitation of the phason modes, with a broad peak from the excitation of the mode with the lowest momentum. Figure \ref{fig: THG-current}(a) predicts five resonances with phason modes. However, we observe that the peak intensity and width seem to decrease with increasing momentum, and hence we conjecture that the contributions from high-momentum modes are too small to be separated from the background. This is consistent with recent experimental and theoretical results on THG in cuprates \cite{kaj_terahertz_2023, fiore_manipulating_2023}, demonstrating that the contribution of the Josephson plasma mode to THG is enhanced for frequencies close to $\omega_J$, corresponding to $\omega_0$ in the model considered here.
We also observe a dominant peak from the resonance at $2\Omega_1 = 2\Delta_\partial^{\mathrm{eq}}$. This contains contributions from quasiparticle excitations as well as the resonance with the Higgs mode.
In Fig. \ref{fig: THG-current}(c) and (d), we choose a two-layer illumination, i.e. $A_1(t) =A_2(t) \neq 0$ and $A_n(t)=0 \;(\forall n\neq 1,2)$. In Fig. \ref{fig: THG-current}(c), both $|j^{ab,(3)}_1|$ and $|j^{ab,(3)}_2|$ show resonant behavior, when $2\Omega_2=2.5$ crosses the energy of the respective Higgs-mode. This is in contrast to Fig. \ref{fig: THG-current}(d), where the boundary Higgs mode has no crossing point with $2\Omega_3=3.34$ and does not exhibit a resonance, whereas we find a resonance with bulk the Higgs mode in the second layer. Remarkably, one can also observe that $|j^{ab,(3)}_1|$ exhibits a small peak when $|j^{ab,(3)}_2|$ is resonantly enhanced and vice versa. This is in accordance with the analytical predictions from Sec. \ref{sec:4}, where we found poles in $\delta \Delta'_n(\omega)$ at $\omega = 2\Delta_\partial^{\mathrm{eq}}$ and $\omega = 2\Delta^{\mathrm{eq}}$ in all the layers. Here, the resonance with the Higgs mode in the first (second) layer leads to enhanced oscillations of the pseudomagnetic field acting on the second (first) layer, which in return results in an amplification of the pseudospin precession. Hence a small peak in the nonlinear current occurs even in layers that are off-resonant. 
\section{Conclusion}
In summary, we have demonstrated that in Josephson coupled layered superconductors the light induced precession of Anderson's pseudospin propagates through the system along the c-axis. Strikingly, this leads to resonant excitation of collective modes in  \textit{all} of the layers, even if the electromagnetic field can only penetrate the surface of the layered structure. We have shown that the ab-polarized light leads to an oscillating Josephson current along the c-axis at resonance with the phason modes, which also contribute to the THG spectrum. Our research, however, was restricted to a simple toy model. Hence, additional research is required to make quantitative predictions for experiments. Firstly, one should incorporate inter- and intralayer Coulomb interactions, which as discussed should open up a gap for the phase mode, as well as quasiparticle tunneling.
The effect of the long-range Coulomb interactions on the optical response has been discussed in the context of the Leggett phase-difference mode in multiband superconductors. Here, the Coulomb interactions do not alter the THG or Raman response, as these oscillations are of intra-unit-cell type \cite{cea_signature_2016,maiti_conservation_2017}. This is in contrast to the phason, which turns into the Josephson plasmon under the influence of the long-range Coulomb interactions, and has been shown experimentally to impact THG \cite{kaj_terahertz_2023}. 
One should also consider the effect of impurities, which have been discussed to affect the THG spectrum \cite{shimano_higgs_2020, Silaev_nonlinear_2019, murotani_nonlinear_2019, tsuji_higgs-mode_2020, li_amplitude_2023,li_collective_2024}. In addition, numerical simulations can be performed away from the static mean-field limit in the attractive Hubbard model through DMFT simulations. This is promising, as Anderson pseudospin precession has already been demonstrated by DMFT simulations in the case of a single layer \cite{tsuji_theory_2015}.

Experimental observation of the Higgs-mode resonance in the undriven layers seems hard, as the nonlinear current is proportional to the applied vector-potential and thus the undriven layers do not contribute to THG. However, the observation of additional peaks from the phase-difference mode as in Fig. \ref{fig: THG-current} would serve as experimental evidence for propagation of a collective mode excitation along the c-axis, induced by in-plane polarized light. For that it is practical to prepare a system with very few layers, resulting in well-separated phase-difference modes. 

\begin{acknowledgments}
N.Z. was supported by the German Academic Scholarship Foundation. K.T. acknowledges support by JSPS KAKENHI (Grant No. JP22K20350 and JP23K17664) and JST PRESTO (Grant No. JPMJPR2256). N.T. acknowledges support by JSPS KAKENHI (Grant No. JP20K03811 and JP24H00191) and JST FOREST (Grant No. JPMJFR2131). 
\end{acknowledgments}
\appendix
\section{\label{Appendix A}Matrix inverse for Eqs. \texorpdfstring{\eqref{eq: e.o.m real part} and \eqref{eq: e.o.m imaginary}}{(20) and (25)}}
In this Appendix, we derive the inverse matrix for the linear equations determining $\delta\Delta_n'$ and $\delta\Delta_n''$ in Eqs. \eqref{eq: e.o.m real part} and \eqref{eq: e.o.m imaginary}. In both cases, the matrices have the following form:
\begin{eqnarray}
    \mathbf{T} = 
    \begin{pmatrix}
        D_1 && 1 && 0&& 0 && ...\\
        B && D && 1 && 0 &&...\\
        0 && 1 && D && 1 && ...\\
        && && ... && && \\
        ... && ... && 1 && D&& B\\
        && ... && 0 && 1 && D_1
        \end{pmatrix}.
\end{eqnarray}
In the case of $D_1=D$ and $B=1$ (which holds if $\Delta_\partial=\Delta$), this would be a symmetric tridiagonal Toeplitz matrix, for which analytic expressions for the inverse exist. Due to the similarity of the problem, we adapt the proof in \cite{hu_analytical_1996} to the present situation. By definition, the elements of the inverse are given by
\begin{align}
    T^{-1}_{ij} = (-1)^{i+j}\frac{A_{ij}}{\det \mathbf{T}}\;,
\end{align}
where $A_{ij}$ is the co-factor, determined by dropping the $i$th row and $j$th column of $\mathbf{T}$ and then evaluating the determinant of the resulting matrix. Let us denote the determinant of a $k$-dimensional symmetric, tridiagonal matrix by $M_k$. If one discards the $i$th row and $j$th column, the matrix obtained can be dissected into three blocks of dimension $(i-1),(j-i), (N-j)$, where $j\geq i$ is assumed (similar for $i>j$). The $(j-i)$ block has the unit diagonal and thus the unit determinant. Developing the determinant with respect to the first column/row yields
\begin{align}
    A_{ij} = \left(D_1 M_{i-2} - BM_{i-3}\right)\left(D_1 M_ {N-j-1} - BM_{N-j-2}\right),
\end{align}
where $M_1 = D, M_0 = 1$ and $M_n = 0$ for $n<0$. The determinant of the whole matrix is given by
\begin{align}
    \det \mathbf{T} = D_1^2M_{N-2} - 2D_1BM_{N-3} + B^2M_{N-4}.
\end{align}
By defining $\kappa$ by
\begin{align}
    D = \begin{cases}
    -2\cosh(\kappa) \; &\mathrm{for} \; D<-2\\
    -2\cos(\kappa) \; &\mathrm{for} \; -2\leq D \leq 2\\
    2\cosh(\kappa)\; &\mathrm{for} \; D>2
    \end{cases},
    \label{eq: diagonal element}
\end{align}
and using the analytical expression for $M_k$ from \cite{hu_analytical_1996}
\begin{equation}
M_k = \sum_{i=0}^{\lfloor k/2 \rfloor}(-1)^{i}\binom{i}{k-i}D^{k-2i},  \label{eq: determinant}  
\end{equation}
where $\lfloor k/2 \rfloor$ denotes the integer part of $\frac{k}{2}$, we find the following:
\footnote{Note that \cite{hu_analytical_1996} defines $D=2\cos(\kappa)$ for $-2\leq D\leq 2$ in \eqref{eq: diagonal element}, however \eqref{eq: determinant} in conjunction with Eq. 1.331 from \cite{gradshteini_tables_2007} yields the correct result with $D=-2\cos(\kappa)$} 
\begin{align}
    M_k = \begin{cases}
    (-1)^k\frac{\sinh((k+1)\kappa)}{\sinh(\kappa)} \; &\mathrm{for} \; D<-2\\
    (-1)^k\frac{\sin((k+1)\kappa)}{\sin(\kappa)} \; &\mathrm{for} \; -2\leq D \leq 2\\
   \frac{\sinh((k+1)\kappa)}{\sinh(\kappa)}\; &\mathrm{for} \; D>2
    \end{cases} .
\end{align}
Hence, the inverse element for $D>2$ is found to be
\small
\begin{align}
    &T^{-1}_{ij} = 
    \frac{D_1 \sinh((i-1)\kappa) - B\sinh((i-2)\kappa)}{\sinh(\kappa)}\times\nonumber
    \\
    &\frac{\left(D_1 \sinh((N-j)\kappa) - B\sinh((N+j-1)\kappa)\right)}{D_1^2\sinh((N-1)\kappa) - 2D_1 B \sinh((N-2)\kappa) + B^2 \sinh((N-3)\kappa))} \label{eq: appendix-inverse}
\end{align}
\normalsize
for $D<-2$ as needed for the amplitude mode. By replacing $\sinh$ with $\sin$, one obtains the inverse matrix elements in the regime $-2\leq D\leq2$ as in the case of the phason excitations.

\subsubsection*{Amplitude mode}
In case of the amplitude mode, we have
\begin{align*}
        &D = \frac{U\zeta(\omega)-1}{J\zeta(\omega)}, \quad D_1=  \frac{U\zeta_\partial(\omega)-1}{J\zeta(\omega)},\quad
   B = \frac{\zeta_\partial(\omega)}{\zeta(\omega)}. \nonumber 
\end{align*}
It can then be shown that the denominator of Eq.~(\ref{eq: appendix-inverse}) can only vanish for $D\leq -2$, which translates to the requirement $\omega\leq \omega_+$, where $\omega_+$ is the solution of $\zeta(\omega_+)=\frac{1}{U\pm 2J}$. Then for $N\gg 1$, we have $N-1\approx N-2\approx N-3\approx N$, leading to the following equation for the poles:
\begin{equation}
    \left(D_1^2-2D_1B+B^2\right)\sinh(\kappa)\sinh(N\kappa)=0,
\end{equation}
which for $\zeta(\omega)\neq 0$ results in Eq.~(\ref{eq: pole-equation-real}).
\subsubsection*{Phase-difference modes}
Similarly, for the phase-difference modes, we have
\begin{align*}
    D = \frac{U\Gamma(\omega)-1}{J\Gamma(\omega)},\quad D_1=  \frac{U\Gamma_\partial(\omega)-1}{J\Gamma(\omega)},\quad
    B= \frac{\Gamma_\partial(\omega)}{\Gamma(\omega)}.
\end{align*}
Some algebra shows that poles are only found for $-2\leq D\leq 2$, translating to the condition that $\omega \leq \omega_{\mathrm{max}}$ where $\omega_{\mathrm{max}}$ solves $\Gamma(\omega_\mathrm{max})=\frac{1}{U-2J}$. As in the case of the amplitude mode, we use $N-1\approx N-2\approx N-3\approx N$, and hence find the following equation for the poles:
\begin{equation}
    \left(D_1^2-2D_1B+B^2\right)\sin(\kappa)\sin(N\kappa)=0.
\end{equation}
This is equivalent to Eq.~(\ref{eq: equation for phase modes}).
\bibliography{multilayer_lib.bib}
\end{document}